\documentclass[aps,
prx,
nofootinbib,
floats,
notitlepage,
superscriptaddress,
]{revtex4-1}

\usepackage{natbib}
\usepackage[utf8]{inputenc}
\usepackage{amsmath, amsthm, amssymb}
\usepackage{enumitem}
\usepackage{graphics,graphicx}
\usepackage{epsfig}
\usepackage{epstopdf}
\usepackage{times}
\usepackage{dcolumn}
\usepackage{url}
\usepackage{color}
\usepackage{stackengine}
\usepackage{tabularx}
\usepackage[all]{xy}
\usepackage{mdwlist}
\usepackage{braket}
\usepackage{xr}
\usepackage{float}

\DeclareMathAlphabet{\altmathcal}{OMS}{cmsy}{m}{n}

\usepackage{longtable}
\usepackage{booktabs}

\usepackage{soul}

\newcommand{\vek}[1]{\boldsymbol{#1}}

\begin{document}

\title{Absence of a resolution limit in in-block nestedness}

\author{Manuel S. Mariani}
\affiliation{Institute of Fundamental and Frontier Sciences, \\University of Electronic Science and Technology of China, Chengdy 610054, PR China.}
\affiliation{URPP Social Networks, Universit\"at Z\"urich, Switzerland}

\author{Mar\'ia J.~Palazzi}
\affiliation{Internet Interdisciplinary Institute (IN3), Universitat Oberta de Catalunya, Barcelona, Catalonia, Spain}

\author{Albert Sol\'e-Ribalta}
\affiliation{URPP Social Networks, Universit\"at Z\"urich, Switzerland}
\affiliation{Internet Interdisciplinary Institute (IN3), Universitat Oberta de Catalunya, Barcelona, Catalonia, Spain}

\author{Javier Borge-Holthoefer}
\affiliation{Internet Interdisciplinary Institute (IN3), Universitat Oberta de Catalunya, Barcelona, Catalonia, Spain}

\author{Claudio J. Tessone}
\affiliation{URPP Social Networks, Universit\"at Z\"urich, Switzerland}

\begin{abstract}
Originally a speculative pattern in ecological networks, the hybrid or compound nested-modular pattern has been confirmed, during the last decade, as a relevant structural arrangement that emerges in a variety of contexts --in  {ecological mutualistic systems} and beyond. This implies shifting the focus from the measurement of nestedness as a global property ({\it macro} level), to the detection of blocks ({\it meso} level) that internally exhibit a high degree of nestedness. Unfortunately, the availability and understanding of the methods to properly detect in-block nested partitions {lie} behind the empirical findings: while a precise quality function of in-block nestedness has been proposed, we {lack an understanding} of its possible inherent constraints. Specifically, while {it is well known that} Newman{-Girvan}'s modularity, and related quality functions, notoriously suffer from a resolution limit that impair their ability to detect small blocks, the potential existence of resolution limits for in-block nestedness is unexplored. Here, we provide empirical, numerical and analytical evidence that {the} in-block nestedness {function} lacks a resolution limit, and thus our capacity to detect correct partitions in networks {via its maximization} depends solely on the accuracy of the {optimization algorithms}. 
\end{abstract}
\maketitle


\section{Introduction}
In-block nestedness has emerged, in the last few years, as an interesting pattern in complex networks. Initially proposed merely as a hypothetical configuration \cite{lewinsohn2006structure}, the idea of hybrid nested-modular structures has gained traction after empirical evidence has shown that such arrangements may play a prominent role in many systems, natural \cite{flores2011statistical,flores2013multi,beckett2013coevolutionary,mello2019insights} and artificial \cite{palazzi2019online}.

From a scientific perspective, the presence of in-block nestedness in real networks is not surprising. Modularity --a mesoscale pattern that considers the organization of nodes in a network as a set of cohesive subgroups \cite{newman2004finding}-- is almost ubiquitous in network structures \cite{zachary1977information,guimera05,eriksen2003modularity,adamic2005political,fortunato2010community}. Nestedness \cite{patterson1986nested,atmar1993measureorder,mariani2019nestedness} --where the interactions of nodes with low low degree are a subset of those with larger degree-- is also a prominent macroscale pattern in ecology \cite{atmar1993measureorder,bascompte2003nested} and beyond \cite{saavedra2011strong,bustos2012dynamics,konig2014nestedness,borge2017emergence}. Both structures emerge as a result of different evolutive pressures and, following this logic, if two such mechanisms are concurrent, then hybrid nested-modular (in-block nested) arrangements are expected to appear.

{And yet, practical approaches} to properly identify in-block nestedness are still scarce. In general, the identification of such compound structures operates sequentially: after the identification of a network partition (usually in terms of modularity \cite{newman2004finding}), nestedness (usually in terms of NODF \cite{almeida2008consistent} {or nestedness temperature~\cite{atmar1993measureorder}}) is computed locally for each block. A possible solution to the limitations of such sequential approach has been recently proposed, as a precise formulation of an appropriate quality function, $\mathcal{I}$ \cite{sole2018revealing}. Formally similar to the popular Newman{-Girvan}'s modularity $Q$, the advantage of $\mathcal{I}$ is that it {can be} {maximized algorithmically}, leading to a proper identification of in-block nested partitions --just like other mesoscale patterns that have recently appeared, e.g. multiple core-periphery structure~\cite{kojaku2017finding}. Thanks to these developments, the presence of in-block nested structures in {very} diverse systems has been confirmed \cite{sole2018revealing,palazzi2019online}. However, the mathematical properties of the in-block nestedness function remain largely unknown.

Here, we examine whether the in-block nestedness function exhibits a resolution limit, similar to the one found for the modularity function~\cite{fortunato2007resolution}.
The existence of such a limit would imply the impossibility to detect interaction blocks smaller than a given scale~\cite{fortunato2007resolution}, potentially making the interpretation of the detected nested blocks ambiguous~\cite{fortunato2016community}.
After some preliminary tests on empirical networks, we show numerically and analytically that in-block nestedness function does not exhibit a resolution limit. Such striking result implies that the identification of the correct in-block nested partition in a network depends only on the accuracy of the heuristics used in the {optimization} process, and not {on} any inherent constraint in the formulation of $\mathcal{I}$ itself.

The rest of the paper is organized as follows. Section~\ref{sec:background} introduces the theoretical concepts examined in this work. Section~\ref{sec:empirical} presents results on some empirical networks which provide {valuable} intuitions for the following sections. Section~\ref{sec:analytic} presents the analytic derivations for in-block nestedness in an idealized family of synthetic networks, proving the absence of a resolution limit. Section~\ref{sec:numeric} {provides} numerical evidence that generalizes the analytical findings. Finally, Section~\ref{sec:conclusion} summarizes the main takeaways and open questions for future research.

\begin{table}[t]
\begin{center}
\begin{tabular}{ |l|l| } 
 \hline
\textbf{Symbol}&\textbf{Variable}  \\ 
 \hline
$s\in\{1,\dots,N\}$ & Node  \\
$N$ & Total number of nodes \\
$(s,t)$ & Edge \\
 $\mathbf{A}=\{A_{st}; A_{st}=1 \text{ iff } (s,t) \text{ is observed}\}$ & Adjacency matrix \\
 $E=\sum_{s,t}A_{st}$ & Total number of edges  \\
$\Omega_s=\{t|A_{st}=1\}$ & Neighborhood of node $s$ \\
 $k_s=\sum_{t}A_{st}$ & Degree of node $s$ \\
$\alpha_{s}$ & Community where node $s$ belongs to\\
$\kappa_{s} = \sum_{t \in \alpha_{s}}A_{st}$ & Internal degree of node $s$\\
$\kappa_{\alpha} = \frac{1}{2}\sum_{s \in \alpha} \kappa_{s}$ & Internal degree of community $\alpha$\\
$O_{st}=\sum_{u}A_{su}\,A_{tu}$ & Common neighbors / overlap \\
 \hline
\end{tabular}
\end{center}
\caption{Basic notation for unipartite networks~\cite{newman2010networks}}
\label{tab:unipartite}
\end{table}

\section{Background on modularity and in-block nestedness}
\label{sec:background}
Heterogeneity is a landmark feature of real complex networks. At the global scale, for example, the distribution of the number of neighbors of a {node} is {broad, with a tail that often follows a power law}. Interestingly, also the mesoscale often presents a similar situation: the distribution of edges is not only globally, but also locally inhomogeneous, with high concentrations of edges {\it within} groups of nodes, and low concentrations {\it between} these groups~\cite{fortunato2010community}. 

Such feature of real networks --community structure-- can be translated to a quantitative criterion. Radicchi \textit{et al.}~\cite{radicchi2004defining} propose the following: a block (also called {\it community}, {\it module}, {\textit{compartment},} or {\it cluster}{depending on the research field~\cite{mariani2019nestedness}}) constitutes a \textit{weak community} if and only if its internal degree $\kappa_{\alpha}$ exceeds its external degree (i.e., the total degree of its nodes by only considering links with nodes that do not belong to the block). Conversely, a block constitutes a \textit{strong community} if and only if, for each of its nodes, the node's internal degree $\kappa_{s}$ is larger than the node's external degree.

Notably, this definition presupposes that a partition of the network is at hand --whereas, in real situations, this is {most} often not the case. This explains why, historically, there have been many efforts to define suitable quality functions, and design associated optimization heuristics, that aim at the identification of good (ideally optimal) partitions. Without doubt, the most popular method in network science is through the maximization of a fitness function called modularity $Q$~\cite{newman2004finding}. Yet, other definitions~\cite{rosvall2008maps} and structures have also attracted the attention of researchers~\cite{borgatti2000models,rombach2017core,kojaku2017finding}.

In this section we focus on two of those functions, that constitute the core of this work (modularity and in-block nestedness~\cite{sole2018revealing}), {emphasizing} their inherent shortcomings, i.e. limitations that are intrinsic to their definition, rather than to the weaknesses of the corresponding optimization strategies. For the sake of simplicity, we only report definitions for unipartite networks, see Table~\ref{tab:unipartite} for the notation used in this work. The extension to bipartite systems is not difficult, but requires more intricate notation, as well as the consideration of the different number of nodes that may compose each network dimension. 


\subsection{Modularity}
\begin{figure}[t]
\centering
\topinset{\bfseries}{\includegraphics[width=.55\textwidth]{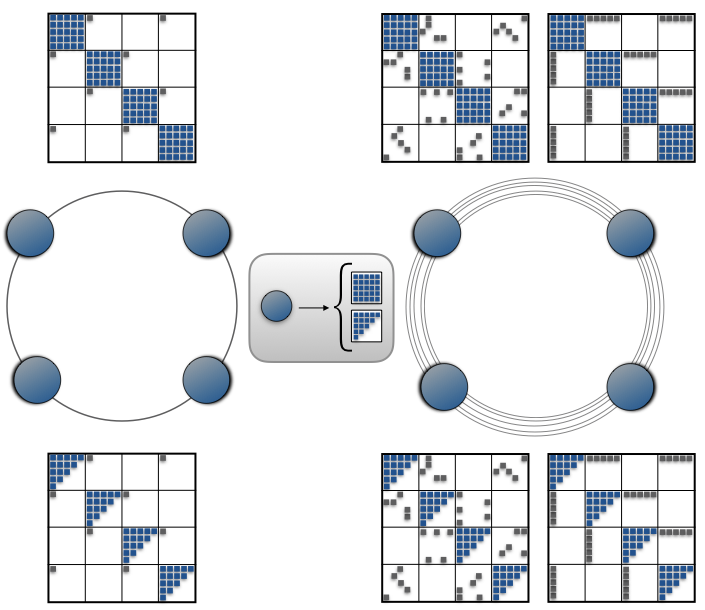}}{0.15in}{0.05in}
\vspace{-.2cm}
\caption{\textbf{Illustration of a ring of weakly-interconnected blocks.} Central row: Representation of a ring of  sub-graphs (blue circles) connected by a single link, $\ell = 1$ (left) and connected through several links $\ell = 5$ (right). The subgraphs, represented as blue circles, can take the form of identical cliques or perfect nested blocks. A matrix representation for the considered cases is shown in top (identical cliques) and bottom (identical perfect nested blocks) rows, respectively.}
\label{fig:fig1}
\end{figure}

One of the most popular methods to identify communities is through the maximization of the modularity $Q$ \cite{fortunato2010community,fortunato2016community}. For a unipartite network, the {modularity function is defined as}~\cite{newman2004finding}
\begin{equation}
Q=\frac{1}{2\,E}\sum_{st}(A_{st}-E_{st})\,\delta(\alpha_s,\alpha_t),
\label{modularity}
\end{equation}
where $E_{st}=k_s\,k_t/(2\,E)$ denotes the expected number of links between nodes $s$ and $t$ under the Chung-Lu configuration model 
\cite{chung2002average,chung2002connected}.
The problem of community detection via modularity optimization is particularly tricky, and has been the subject of discussion in various disciplines. This is an NP complete problem \cite{garey1979computers}, which explains why several methods have been proposed to reduce the complexity of the task \cite{danon2005comparing,peixoto2013parsimonious,sobolevsky2014general}.
However, parallel to the constraints of the algorithmic strategies, the formulation of $Q$ has an inherent limitation itself, which impedes its optimization to detect blocks that are smaller than a given size. Intuitively, for the modularity function, this limit can be understood in a toy unipartite network formed by a set of cliques placed on a ring, where each pair of adjacent cliques is connected by a single inter-clique link, see Figure~\ref{fig:fig1} (top left and middle left panels). This is the most modular connected network \cite{fortunato2007resolution}. In this setting, one can show that the modularity has a scale detection problem. Even if the network has more cliques than $B\geq \sqrt{E}$, the modularity function will still favor partitions where $B$ blocks are detected. This somehow imposes a detection scale which can be intuitively understood by noticing that the expected number of edges between two blocks $\alpha$ and $\beta$ is, approximately, $E_{\alpha\beta}=k_\alpha\,k_\beta/(2\,E)$, where $k_\alpha=\sum_{s\in\alpha}k_s$ denotes the total degree of block $\alpha$. When both $k_\alpha$ and $k_\beta$ are of order $\sqrt{E}$ or smaller, $E_{\alpha\beta}$ becomes of order one or smaller, meaning that even a single link between blocks $\alpha$ and $\beta$ is interpreted by the modularity function as a non-random connection, thereby favoring their merging into a single block~\cite{fortunato2016community}.


In the maximally-modular network above, an alternative demonstration of the resolution limit can be obtained by comparing the modularity of the correct partition of the nodes into cliques, $Q_{single}$, against the modularity of the (wrong) partition obtained by merging pairs of adjacent cliques, $Q_{pairs}$. It turns out that $\Delta Q:=Q_{single}-Q_{pairs}>0$ if and only if $N<\sqrt{E}$. If we gradually increase $N$ by adding new cliques, as soon as $N$ becomes larger than the modularity's intrinsic scale $\sqrt{E}$, the modularity of the wrong partition, $Q_{pairs}$, exceeds the modularity of the correct partition, $Q_{single}$ ($\Delta Q<0$). Alternative examples can be drawn to further prove the modularity's resolution limit in various scenarios~\cite{fortunato2007resolution}.

\subsection{In-block nestedness}
Among mesoscale quality functions other than modularity, in-block nestedness corresponds to a hybrid or combined pattern in which nested-modular arrangements appear. To be precise, an in-block nested network presents an overall {compartmentalized} organization, where blocks present a nested connectivity within.
Naturally, it follows that the in-block nestedness quality function $\mathcal{I}$ inherits aspects from nestedness measurement (in particular, from the NODF descriptor \cite{almeida2008consistent,ulrich2009consumer}), as well as ingredients from modularity.  
For a unipartite network, the in-block nestedness function $\mathcal{I}$ is defined as~\cite{sole2018revealing}
\begin{equation}
\mathcal{I}=\frac{2}{N}\sum_{st}\frac{O_{st}-\braket{O_{st}}}{k_t\,(C_s-1)}\,\Theta(k_s-k_t)\,\delta(\alpha_s,\alpha_t),
\label{ibn1}
\end{equation}
where each node can only belong to one block $\alpha$, $C_\alpha$ denotes the number of nodes that belong to block $\alpha$, $O_{st}$ the number of shared neighbours between nodes $s$ and $t$ (i.e. overlap), $\Theta$ is the Heaviside function and $\delta$ is the Kronecker delta.
For subsequent analytic developments, it is convenient to rewrite the previous expression for $\mathcal{I}$ as sum over the network's blocks:
\begin{equation}
\mathcal{I}=\sum_{\alpha=1}^{B}\mathcal{N}_{\alpha}
\label{ibn_block}
\end{equation}
where {$B$ denotes the total number of blocks and}
\begin{equation}
   \mathcal{N}_{\alpha}:= \frac{2}{N}\frac{1}{C_{\alpha}-1}\sum_{s,t\in\alpha}\Biggl(\frac{O_{st}}{k_t}-\frac{k_s}{N} \Biggr)\,\Theta(k_s-k_t).
\label{component}
\end{equation}
{can be interpreted as the level of block $\alpha$'s internal nestedness.}

Measuring the level of in-block nestedness of a given network requires the optimization of the in-block nestedness function, which is --again-- an NP problem. Leaving aside computational aspects, resolution limits can arise when optimizing a quality function different than modularity~\cite{traag2011narrow}. For instance, recent works have introduced and examined a quality function that assumes that each block has a core-periphery internal structure~\cite{kojaku2017finding,kojaku2018core}. While this quality function can detect multiple core-periphery structures in a network, it inherits from the modularity function a similar resolution limit~\cite{kojaku2018core}, which has motivated the introduction of a multiscale variant of the original algorithm~\cite{kojaku2019multiscale}.

Turning to in-block nestedness, the function defined by Eq.~\eqref{ibn1} is substantially different than the modularity function, because it is based on the overlaps between nodes that belong to the same cluster, and not on link density. This suggests that the resolution limit of this function may have a radically different behavior than the modularity's one.
Examining this conjecture is the main goal of the rest of the paper.


\section{Empirical insights: preliminary intuitions on $Q$ and $\mathcal{I}$ resolution limit}
\label{sec:empirical}

To test the absence (or presence) of a resolution limit for the in-block nestedness, we first perform an exploration with empirical data, following an approach similar to the one in  \cite{fortunato2007resolution}. Specifically, for each network, the quality function of interest is optimized by means of the same optimization strategy (extremal optimization algorithm \cite{duch2005community}, {in our case}. See Appendix~\ref{sec:optimization}). Then, all the links between the detected blocks are removed, and the optimization algorithm is applied again to the resulting blocks. With two partitions at hand, we compute the Jaccard index to measure how similar they are. We iterate this procedure --remove links between communities, optimize quality function--, until the Jaccard index $J$ between consecutive partition vectors is 1, i.e. the algorithm is no longer able to split the current partition into one with higher score. 

This general scheme is applied separately for modularity $Q$ and in-block nestedness $\altmathcal{I}$ over a set of 82 real networks, from two different domains: ecological in most cases \cite{wol}, with some collaboration networks taken from socio-technological systems~\cite{Github,palazzi2019online} (see Appendix \ref{sec:data}  for details). We have restricted the size of these networks in the range $[50,10^3]$ nodes.

The idea behind this approach is to get a first intuition on whether a resolution limit for in-block nestedness exists, or not, and how severe it is --if it does exist--, when compared to the resolution limit of modularity. If the quality function lacks a resolution limit (and assuming that the heuristics can reach the optimal partition), one should expect that after the initial optimization step, the algorithm should not be able to further split the detected blocks into smaller ones.

\begin{figure}[b]
\centering
\def\stackalignment{l}
\topinset{\bfseries(a)}{\includegraphics[width=.45\textwidth]{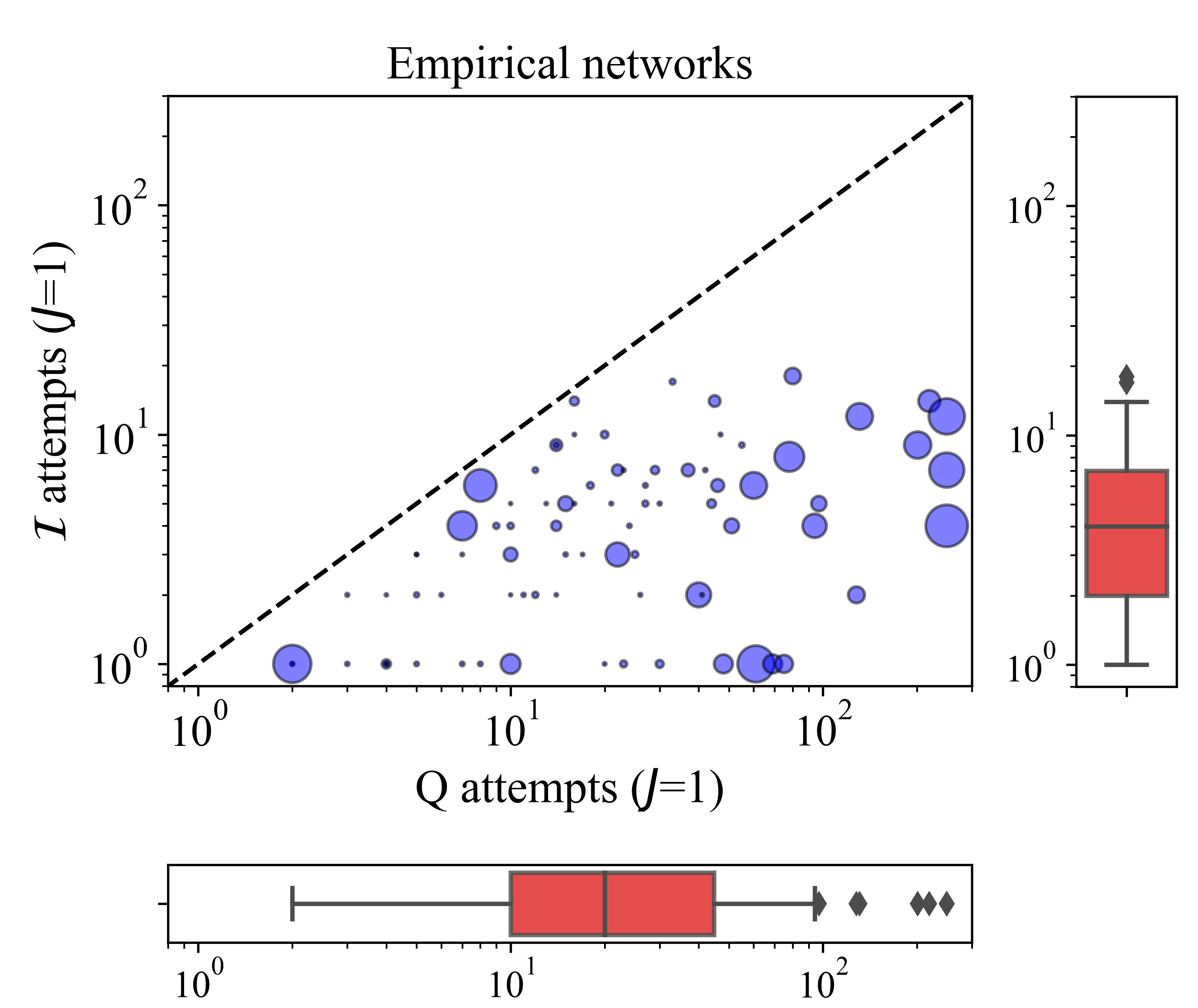}}{0.3in}{0.5in}
\topinset{\bfseries(b)}{\includegraphics[width=.45\textwidth]{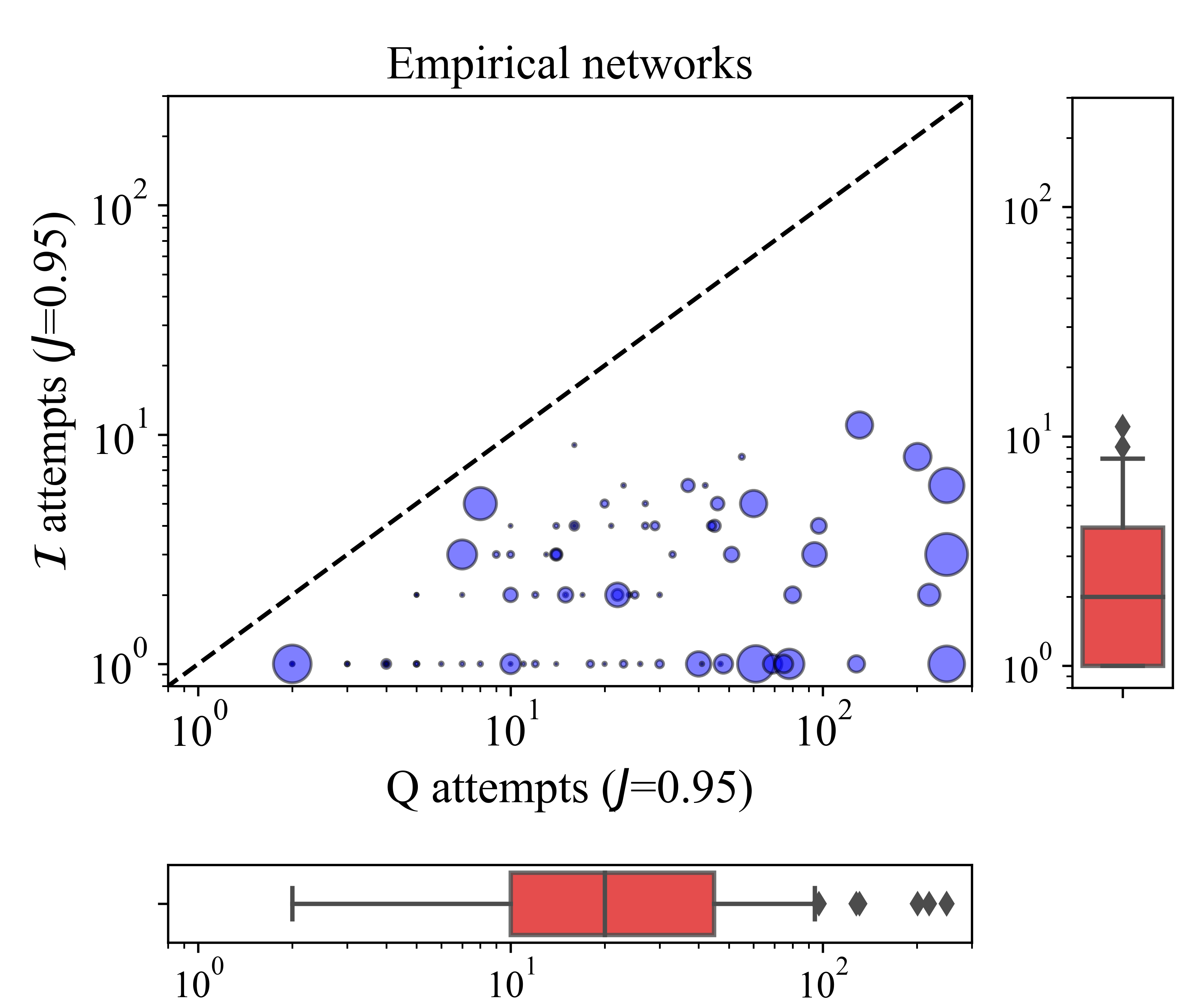}}{0.3in}{0.5in}
\vspace{-.2cm}
\caption{\textbf{Comparing the resolution limit of modularity and in-block nestedness in empirical data.} Scatter plots of the number of attempts needed to reach a Jaccard index $J=1$ (left) and $J=0.95$ (right) for modularity and in-block nestedness. Marginal box plots show the distribution of the number of attempts needed for each network. The size of the points in the scatter plot is proportional to the total number of nodes of each network.}
\label{fig:fig3}
\end{figure}

The result of this experiment is summarized in Figure~\ref{fig:fig3} (left panel) which shows a scatter plot of the number of attempts needed to reach $J=1$ after optimizing in-block nestedness, plotted against the corresponding number of attempts to reach $J=1$ for modularity, for each network. The size of the points in the scatter plot is proportional to the size of each network. To ease comparison, the number of attempts for $Q$ and $\altmathcal{I}$ have been plotted in the same scale (log-log), the function $y=x$ is plotted as a dashed black curve as a visual aid. Marginal box plots show the distribution of the number of attempts needed for each network, for both $Q$ and $\altmathcal{I}$. Without exception, the number of iterations needed to reach the stopping condition is {substantially} longer for modularity. 

Taken strictly, this result can be interpreted as informal evidence of a milder effect of the resolution limit for in-block nestedness (compared to $Q$). At the same time, this result is not a formal proof that the resolution limit is entirely absent: if the resolution limit is absent, the additional optimization steps could be due to the fact that the extremal optimization algorithm is unable to reach the optimal partition in each step. Relaxing the conditions for the stopping criteria, e.g.  $J \geq \tau$, with $\tau \in [0.95, 0.99]$, strengthens this informal evidence: the number of attempts needed to reach $J \geq \tau$ for $\altmathcal{I}$ drops to 1 for several networks, while the number of attempts for $Q$ remains large in most cases: see Figure~\ref{fig:fig3} (right panel), which shows this for $\tau = 0.95$.

\section{Absence of resolution limit in $\mathcal{I}$: analytic approach}
\label{sec:analytic}

\begin{figure}[t]
\centering
\def\stackalignment{l}
\topinset{\bfseries(a)}{\includegraphics[width=0.45\textwidth]{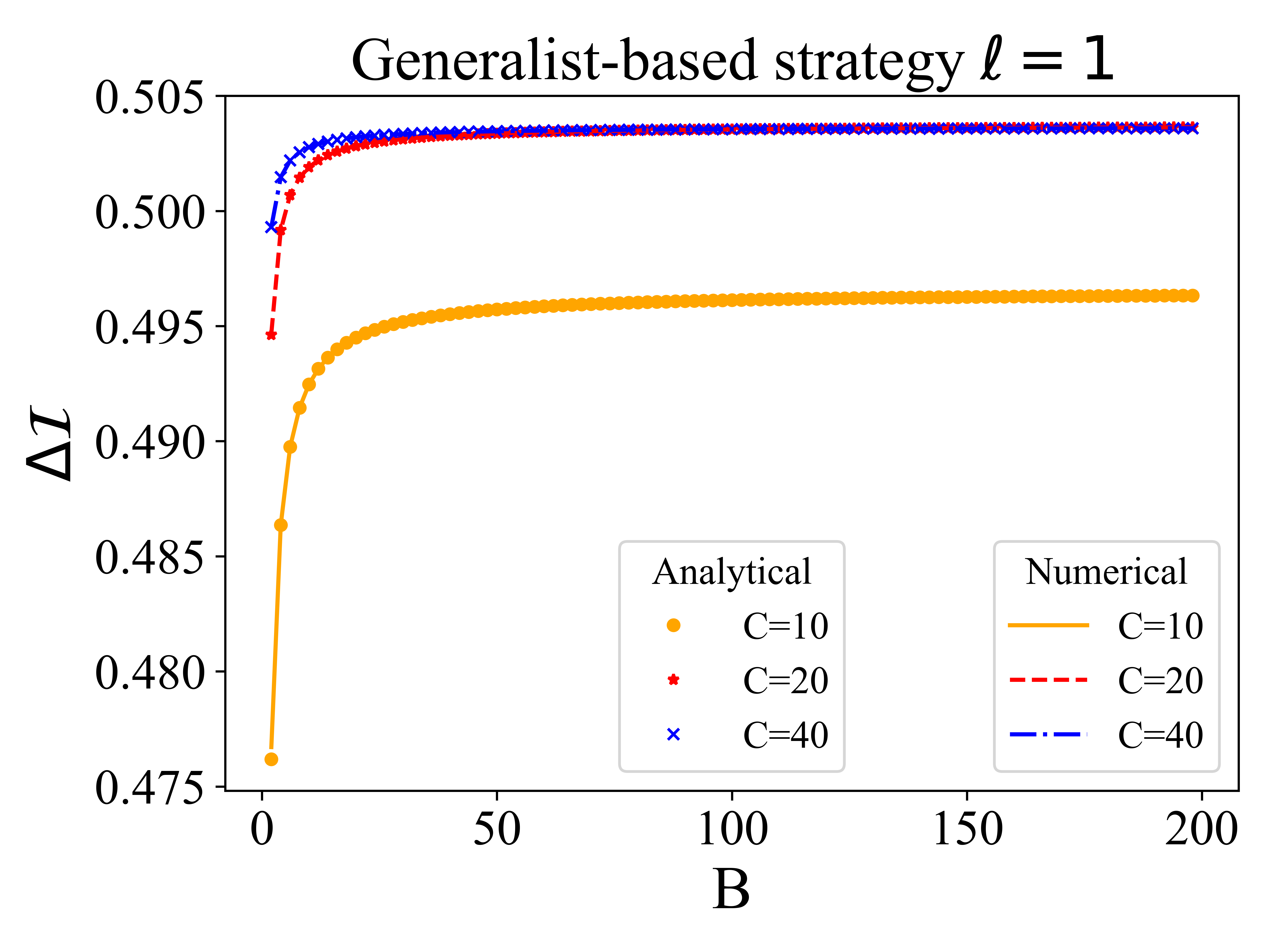}}{0.15in}{0.05in}
\topinset{\bfseries(b)}{\includegraphics[width=0.45\textwidth]{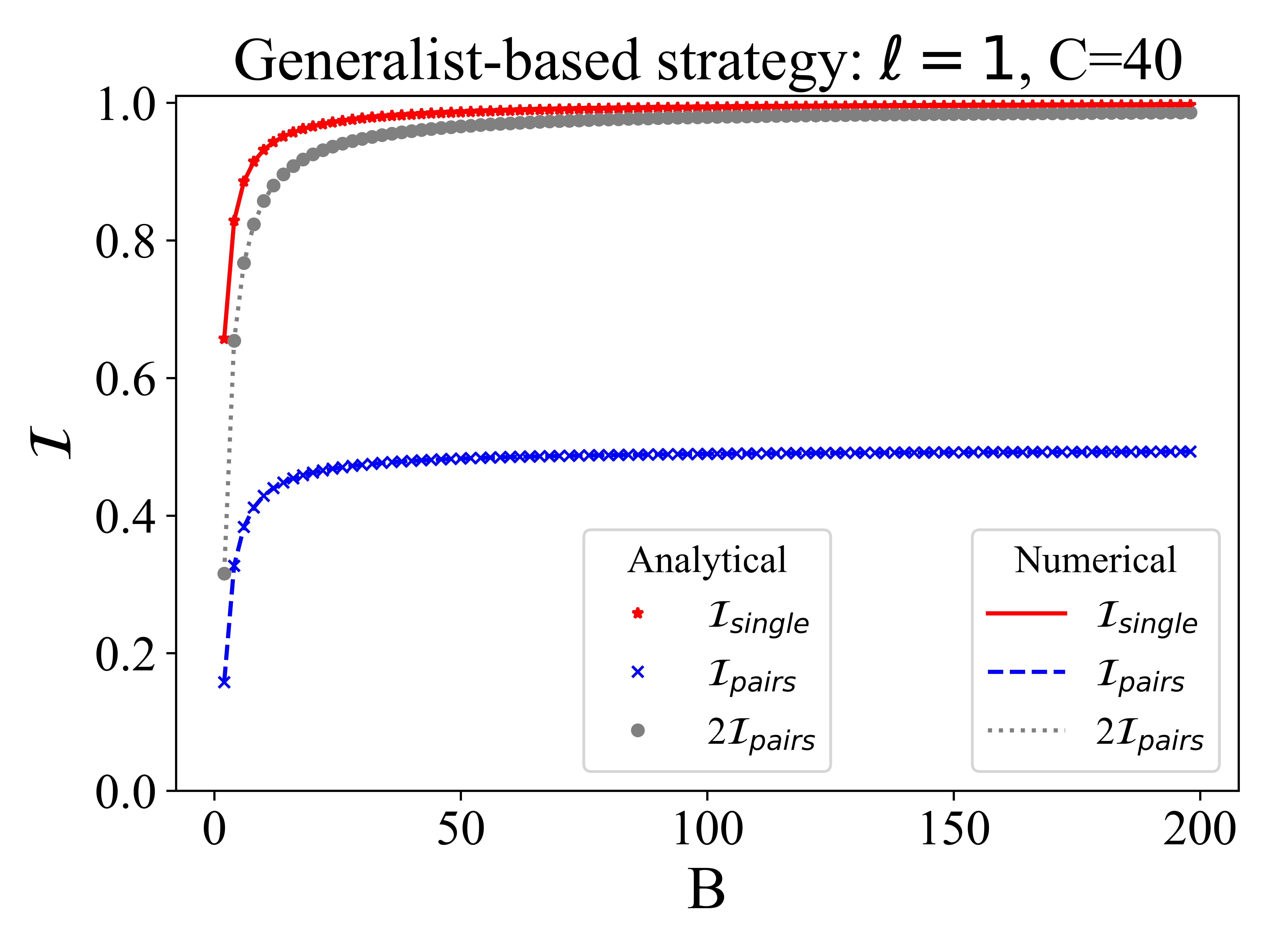}}{0.15in}{0.05in}
\vspace{-.2cm}\caption{\textbf{Analytical and numerical agreement.} Left panel reports on the perfect match between Eq.~\eqref{final} (symbols) and the actual calculation performed on synthetic graphs (lines). Similarly, the right panel shows the agreement between analytical and numerical results for Eqs.~\eqref{Isingle} and \eqref{Ipairs}. Gray symbols and dotted line in right panel confirms Eq.~\eqref{limit}.}
\label{fig:anal_nume}
\end{figure}

In this Section, we aim to provide an analytic explanation for the previous empirical intuitions, in an idealized family of synthetic networks. For the sake of analytic tractability, we consider a ring of interconnected blocks of equal size $C$, where each block has internally a stepwise structure. That is, the degrees of subsequent  rows (columns) of the adjacency matrix differ by one (see Fig.~\ref{fig:fig1}, bottom-left panel). Additionally, contiguous blocks are interconnected by $\ell=1$ link that connects the two generalists of each block --in total, there are $B+1$ inter-block links that connect the $B$ generalists. Our strategy to perform the calculation is to first compute in-block nestedness $\mathcal{I}_0$ of a perfectly in-block nested network composed of $B$ disconnected blocks, and then add to $\mathcal{I}_0$ the terms due to the interactions between the hubs. To compute $\mathcal{I}_0$, it is sufficient to derive the nestedness of a single stepwise block, see Appendix \ref{derivation1}.
In the case of equally-sized blocks, we obtain
\begin{equation}
   \mathcal{I}_0=1-\frac{2}{3\,N}-\frac{2}{3\,B}.
   \label{unperturbed}
\end{equation}
From this Equation, we can obtain the in-block nestedness of the ring by adding the contribution to $\mathcal{I}$ from the inter-block links that connect the hubs, see Appendix \ref{derivation2}. 
We obtain
\begin{equation}
  \mathcal{I}_{single}=1-\frac{2}{3\,N}-\frac{2}{3\,B}-\frac{4\,B}{N^2}.
  \label{Isingle}
\end{equation}
To study the possible existence of a resolution limit of $\mathcal{I}$, we need to compare the in-block nestedness of the correct partition, $\mathcal{I}_{single}$, with the in-block nestedness $\mathcal{I}_{pairs}$ of a partition where pairs of contiguous ``true'' blocks, $\alpha_i$ and $\alpha_{i+1}$ ($i=1,\dots,B$), are merged and assigned to the same block, $\alpha_{i,i+1}=\alpha_i\cup \alpha_{i+1}$. 
The in-block nestedness of the wrong partition, $\mathcal{I}_{pairs}$, can be calculated by adding up the contributions from pairs of nodes that belong to the same true block, and those from pairs of nodes that belong to different true blocks. This results in
\begin{equation}
    \mathcal{I}_{pairs}=\frac{C-1}{2\,C-1}\,\mathcal{I}^{single}+\frac{2}{C\,(2\,C-1)}\,\Biggl( H_{C-1}-\frac{g(C)}{3\,N} \Biggr)
    \label{Ipairs}
\end{equation}
where $H_{C-1}:=\sum_{t=1}^{C-1}t^{-1}$ denotes the $C-1$th harmonic number, and we defined the polynomial function $g(C):=(C-1)\,(C^2+C+6)$. Putting together Eqs.~\ref{Isingle} and~\ref{Ipairs}, we obtain
\begin{equation}
   \Delta\mathcal{I}=\mathcal{I}_{single}-\mathcal{I}_{pairs}=\frac{C}{2\,C-1}\,\mathcal{I}_{single}-\frac{2}{C\,(2\,C-1)}\,\Biggl( H_{C-1}-\frac{g(C)}{3\,N} \Biggr).
   \label{final}
\end{equation}
We refer to Appendix \ref{derivation2} for the full derivation of this equation. Also, numerical results in Figure~\ref{fig:anal_nume} show the perfect matching between the analytical insights in Eq.~\eqref{final} (left panel), and Eqs.~\eqref{Isingle},\eqref{Ipairs} and \eqref{limit} (right panel).

For a fixed $C\gg 1$ value, in the limit $N\to\infty$ (or equivalently, $B\to\infty$), we obtain 
\begin{equation}
\mathcal{I}^{pairs}\to \mathcal{I}^{single}/2
\label{limit}
\end{equation}
confirming the numerical intutions in~\cite{sole2018revealing}, and in accordance with the right panel of Figure~\ref{fig:anal_nume}. This implies that no matter how large the network is, the in-block nestedness of the partition with pairwise-merged blocks remains significantly smaller than the in-block nestedness of the partition with the original blocks. The same holds true for small values of $C$, because the second term in the r.h.s. of Eq.~\eqref{final} tends to be substantially smaller than the first term.
The reason is that the contribution from the null model is negligible compared to the penalty due to the merging of two blocks into a single one. Therefore, in this idealized example, the penalization for larger blocks in the in-block nestedness function prevents the resolution limit, allowing the in-block nestedness function of the partition composed of the individual blocks to stay always larger than the in-block nestedness of the partition composed of pairwise-merged blocks.

\section{Generalizing the absence of resolution limit in $\mathcal{I}$: numerical approach on benchmark graphs}
\label{sec:numeric}

Supported by the excellent agreement between analytical and numerical results in Figure~\ref{fig:anal_nume}, we now carry out a numerical validation considering less idealized scenarios. We do so examining numerically whether the in-block nestedness function presents a resolution limit or not, in scenarios beyond $\ell = 1$ where modularity does. To this end, we analyze benchmark networks along the lines of Figure~\ref{fig:fig1} (middle-right and bottom-right panels), that is, building unipartite synthetic networks, composed of a growing ring of blocks that internally exhibit a nested structure. We study a wide range of these networks, modifying the number of blocks $B$ that conform the ring, and the number of inter-block links $\ell$. We start with a network composed of $B=3$ (perfectly nested) stepwise blocks connected as a ring, and then consider a growing number of blocks (up to $B=200$). Regarding the inter-block connectivity $\ell$, we start with $\ell = 1$, which corresponds to the analytical calculations above, up to $\ell = C(C - 1)/2$ which corresponds to maximum possible connectivity between contiguous blocks. Details on the generation of the internal nested structure of the blocks, and how it determines the internal block density, is available in Appendix \ref{sec:toy}.

We carry out the numerical validation in two flavors: in one of them, we consider a \textit{random strategy}, where the blocks are connected by adding a link between two randomly selected nodes from each block. For this case, we report results averaged over 25 realizations. In the other, the addition of inter-block links ($\ell \geq 1$) is deterministic, connecting the most-generalist available nodes in each pair of adjacent communities (\textit{generalist-based strategy}). Note that, strictly speaking, this latter strategy is the logical generalization of our analytical results (where a single link was laid between adjacent blocks, connecting the generalist nodes in them). 

\begin{figure}[t]
\centering
\def\stackalignment{l}
\topinset{\bfseries(a)}{\includegraphics[width=0.95\textwidth]{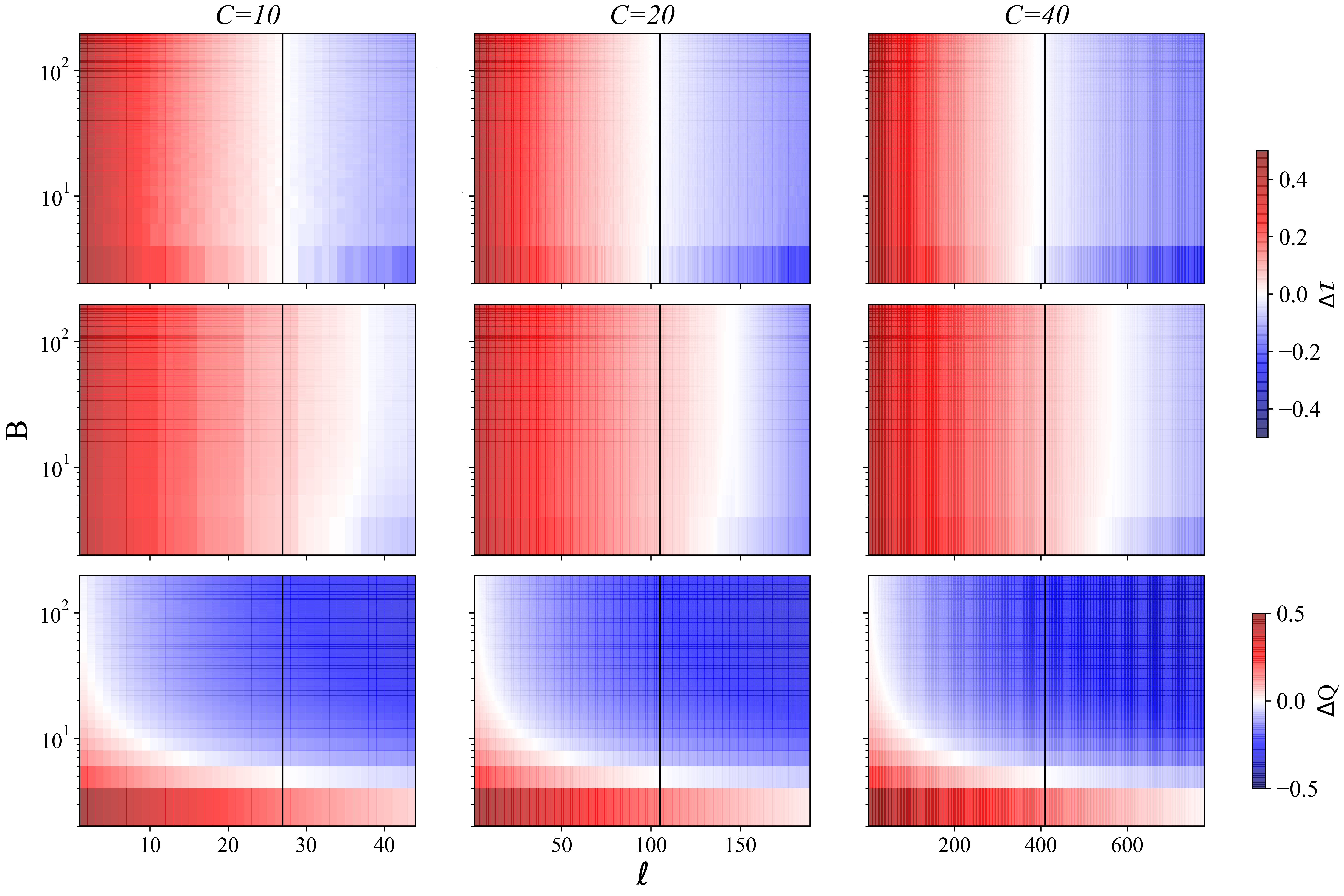}}{0.25in}{0.46in}\topinset{\bfseries(b)}{}{-4.22in}{-4.29in}\topinset{\bfseries(c)}{}{-4.22in}{-2.29in}
\topinset{\bfseries(d)}{}{-2.865in}{-6.28in}\topinset{\bfseries(e)}{}{-2.865in}{-4.315in}\topinset{\bfseries(f)}{}{-2.865in}{-2.312in}
\topinset{\bfseries(g)}{}{-1.511in}{-6.3139in}\topinset{\bfseries(h)}{}{-1.511in}{-4.35in}\topinset{\bfseries(i)}{}{-1.511in}{-2.339 in}
\vspace{-.2cm}\caption{\textbf{Resolution limit in random- and generalist-connected rings of nested blocks.} The panels represent three-dimensional plots in the parameter space $(B,\ell)$ showing, in the $z$-axis (color code), the values of $\Delta\altmathcal{I}$ (top and middle panels) and $\Delta Q$ (bottom panel). In the random linking strategy (top and bottom), results are averaged over 25 different realizations. The solid black line indicates the transition from weak communites to no communities, as defined by Radicchi {\it et al.} \cite{radicchi2004defining}.}
\label{fig:metrics_1}
\end{figure}

For both strategies, we compare numerically the in-block nestedness of the ground-truth partition, $\mathcal{I}_{single}$, against the in-block nestedness of the wrong partition obtained by considering pairs of adjacent blocks as a single block, $\mathcal{I}_{pairs}$. If the in-block nestedness has a resolution limit beyond the scenario presented in the previous section, then for some value of $B$ we would observe a crossover from $\Delta\mathcal{I}:=\mathcal{I}_{single} - \mathcal{I}_{pairs} > 0$ to $\Delta\mathcal{I} < 0$, as indeed happens with $Q$.
All these results are shown in Figure~\ref{fig:metrics_1}, where top and middle rows present the results for $\Delta\mathcal{I}$ in the random and generalist strategies, respectively. The bottom row, conversely, corresponds to the results for $\Delta Q$ for {\it both} strategies, since they present --quite surprisingly-- identical behavior. For the sake of clarity, a black vertical line is drawn in each panel marking the weak community criterion. Beyond this limit, no recognizable block structure is available, and therefore it becomes irrelevant whether a given quality function identifies a ``correct'' block or not. Each column of the figure corresponds to different block sizes $C$. 


For the random strategy, each point in the parameter space $(B, \ell)$ of the panels in Figure~\ref{fig:metrics_1} reports the average value of $\Delta \altmathcal{I}$ (top row), and $\Delta Q$ (bottom row), for 25 different realizations. 
There are at least three remarkable lessons from Figure~\ref{fig:metrics_1}, equally valid for the adopted linking strategies. 
First, only $Q$ shows the existence of a resolution limit consistently --no matter the number of inter-block links $\ell$, it is always possible to find a large-enough number of blocks such that the resolution limit appears, i.e. $Q_{pairs}$ is larger than $Q_{single}$. Second (a consequence of the first), the appearance of the resolution limit for $Q$ is independent of the criterion of weak community: the crossover to $\Delta Q<0$ can occur anywhere in the $\ell$ {spectrum}, and {it depends} on $B$ only (i.e., on network size, in line with the analytic results in~\cite{fortunato2007resolution}). Of course, increasing $\ell$ reduces the amount of blocks $B$ needed to reach the crossover (note the logarithmic scale on the $B$ axis). 
Finally, the robustness of the single block as the best partitioning scheme for in-block nestedness (i.e. $\Delta \altmathcal{I} > 0$ is remarkably high: note that $\altmathcal{I}_{single}$ remains systematically larger than $\altmathcal{I}_{pairs}$ until $\ell$ has almost reached the weak community criterion. In other words, $\altmathcal{I}$ identifies the correct block-by-block structure up to the point where such partition (or any other one) becomes unrecognizable.

The only relevant difference between the random (top) and the generalist (middle) linking strategies is related to $\ell$: the area of the parameter space where the in-block nestedness cannot detect the correct block partition ($\Delta\mathcal{I} < 0$) is substantially smaller in the generalist strategy, compared to the same area under the random strategy. This indicates that when inter-block connections are preferentially established by local hubs (or generalists), in-block nestedness can detect blocks of locally nested interactions even when these blocks are not communities in the traditional sense. Other than this important remark, the previous conclusion holds: $\altmathcal{I}$ does not show a dependency on $B$ (and thus on $N$) by which its ability to detect the right partition is affected, and thus $\altmathcal{I}$ appears to lack a size-related resolution limit.

\section{Conclusions}
\label{sec:conclusion}


During the last few years, in-block nestedness has become a relevant structural arrangement in complex networks and a precise formulation of an appropriate quality function to detect in-block nested patterns has been recently introduced. Nonetheless, the possible inherent constraints of this quality function are still largely unknown. Particularly, the potential existence of a resolution limit for in-block nestedness --similar to the one found for modularity-- remains unexplored.

In this work, we have verified whether the in-block nestedness function exhibits a  modularity-like resolution limit, i.e., the inability to identify blocks smaller than a certain scale. We have approached the question of in-block nestedness' resolution limit as a three-step process. First, we have performed an informal test on empirical networks, to assess the extent to which a network can be recursively split into smaller and smaller blocks, which is an indication of the existence of a resolution limit \cite{fortunato2007resolution}. From there, upon the intuition that in-block nestedness lacks a resolution limit (or, at least, it is less severe than $Q$'s), we provide a formal {proof} that $\altmathcal{I}$ {does not have a resolution limit}, at least in a specific setting --that in which different blocks are connected by a single link. Finally, we have numerically generalized and confirmed the analytical argument, exhaustively studying a large parameter space with varying network size and inter-block connectivity.

A limitation of our study is that we have focused on the resolution limit that characterizes the modularity function~\cite{fortunato2007resolution}. Our results do not rule out the possibility that in-block nestedness might exhibit different kinds of biases in favor of different properties, e.g. specific intra-block densities, or block relative size distribution. Additional studies on alternative sources of biases of existing quality functions for network analysis are of utmost importance in order to accurately understand the architecture of ecological and socioeconomic systems.

\section*{Acknowledgements}

MSM and CJT acknowledge financial support from the URPP Social Networks at the University of Zurich, and the Swiss National Science Foundation (Grant No. 200021-182659).
MSM acknowledges financial support from the Science Strength Promotion Program of the UESTC, and the UESTC professor research start-up (Grant No. ZYGX2018KYQD215). M.J.P, A.S-R. and J.B-H. acknowledge the support of the Spanish MICINN project PGC2018-096999-A-I00. M.J.P. acknowledges as well the support of a doctoral grant from the Universitat Oberta de Catalunya (UOC).

\appendix

\section{Empirical datasets}
\label{sec:data}
The empirical ecological networks analyzed here represent bipartite mutualistic and competitive systems, including macroscopic and microscopic environments. Network data can be downloaded from \cite{wol} in different formats, and can be filtered depending on the type of interaction of the system (e.g. plant-pollinator, host-parasite) and the type of data, e.g. binary or weighted. In this work, we have analyzed a total of $52$ of these networks, all of them in their binary form. Thus, this kind of networks are represented as a rectangular $N \times M$ matrix, where rows and columns refer to interacting species. An entry in the matrix $a_{ij}= 1$ if species $i$ of one guild interacts with a species $j$ of the other guild at least once, and 0 otherwise.

On the other hand, for the collaboration networks we collected data from open source software projects dataset through GitHub \cite{Github}, a social coding platform that provides source code management and collaboration features. Similar to ecological networks above, for each project ($30$ in total) we build a bipartite unweighted network as a rectangular $N \times M$ matrix, where rows and columns refer to the contributors and source files of each open source software project, respectively. An entry in the matrix $a_{ij}= 1$ if a contributor $i$ have edited a file $j$ at least once, and 0 otherwise. More details on this dataset can be found in \cite{palazzi2019online}. The dataset with the OSS projects is available at \url{http://cosin3.rdi.uoc.edu}, under the Resources section.

\section{Optimization algorithm}
\label{sec:optimization}
As mentioned in the main text, we have employed the extremal optimization algorithm to maximize the modularity and in-block nestedness quality functions. This algorithm was adapted for modularity optimization by Duch and Arenas \cite{duch2005community}. Notably, it offers a good trade-off between accuracy and computational speed. Additionally, the ``simplicity'' of the algorithm, based on the optimization of local variables, facilitates its adaptation to the case of the in-block nestedness quality function.

The algorithm proceeds as follows: starting from a random partition of a network into two groups with the same number of nodes, at each step, a local fitness measure for each node is calculated by dividing the local fitness of the node by its degree. With some probability, the node with the lowest fitness is moved to the other partition. Each movement implies a change in the partition, and a recalculation of the fitness is performed. The process is then repeated until the global fitness score can no longer be improved. Once such bipartition is at hand, each subgraph is considered as a graph on its own, and the procedure is repeated recursively for each one, as long as the fitness function increases with each subsequent partition. 

The corresponding software codes for modularity and in-block nestedness optimization (for uni- and bipartite cases), can be downloaded from the web page \url{http://cosin3.rdi.uoc.edu/}, under the Resources section.

\section{Benchmark graph model}
\label{sec:toy}

The internal nested structure of the blocks is generated by defining a separatrix line that divides the filled and empty regions of the adjacency matrix. 
Following~\cite{sole2018revealing,palazzi2019macro}, we partition the $[0,1]\times[0,1]$ bi-dimensional plane into $N\times N$ cells, and then we only add a link to the cells whose center lies above the separatrix\footnote{Note that this procedure generate unipartite networks with self-loops, which might be an unrealistic trait for some empirical networks. Nevertheless, our results are robust with respect to the removal of self-loops.}. 
We define the separatrix as
\begin{equation}
  f(x)=1-(1-x^{1/\xi})^{\xi}
\end{equation}
where $x\in[0,1]$. Parameter $\xi\in(0,\infty)$controls the slimness of the nested structure and, as a consequence, the internal density of the blocks. In the following, we shall often set $\xi=1$, which corresponds to a stepwise block where given two consecutive rows $i$ and $i+1$, $k_{i+1}-k_{i}=1$. The corresponding software codes, to generate nested, modular, and in-block nested networks, for uni- and bipartite cases, can be downloaded from the web page of the group  \url{http://cosin3.rdi.uoc.edu/}, under the Resources section. 

\section{Proving the absence of a resolution limit in a ring of weakly-interconnected blocks}

In this Section, we derive the analytic results presented in the main text, namely, the in-block nestedness of a set of disconnected stepwise blocks (Eq.~\eqref{unperturbed}, derivation in Appendix~\ref{derivation1}) and the results needed to prove the absence of a resolution limit (Eqs.~\eqref{Isingle}--\eqref{final}, derivation in Appendix~\ref{derivation2}).

\subsection{Derivation of the in-block nestedness of a set of disconnected stepwise blocks}
\label{derivation1}

As each block has an internally nested structure, by definition, $O_{st}=k_t$ if $k_t<k_s$.
Therefore, Eq.~\eqref{component} becomes
\begin{equation}
    \mathcal{N}_{\alpha}=\frac{2\, f(\vek{k}^{(\alpha)})}{N\,(C_{\alpha}-1)},
    \label{component2}
\end{equation}
where $f(\vek{k}^{(\alpha)})$ denotes the following function of the vector $\vek{k}^{(\alpha)}$ composed of the degrees of the nodes that belong to node $\alpha$:
\begin{equation}
    f(\vek{k}^{(\alpha)}):=\sum_{s\in\alpha}\Biggl(1-\frac{k_s}{N}\Biggr)\,\sum_{t\in\alpha}\Theta(k_s-k_t).
\end{equation}
Note that in general, the function $f(\vek{k}^{(\alpha)})$ depends on the perfectly-nested block's internal shape or, equivalently, on the density of the perfectly-nested block. 
The factor $\sum_{t\in\alpha}\Theta(k_s-k_t)$ represents the number of nodes with degree strictly smaller than $k_s$. As we are considering stepwise perfectly nested networks, we have $\sum_{t\in\alpha}\Theta(k_s-k_t)=k_s-1$.
Hence, after rearranging some terms, 
\begin{equation}
    f(\vek{k}^{(\alpha)}):=\Biggl(1+\frac{1}{N}\Biggr)\sum_{s\in\alpha}k_s-\frac{1}{N}\sum_{s\in\alpha}k_s^2-C_{\alpha}.
    \label{intermediate}
\end{equation}
For stepwise perfectly-nested networks, the following identities hold:
\begin{equation}
\begin{split}
&\sum_{s\in\alpha}k_s=\sum_{s=1}^{C_{\alpha}}k_s=\sum_{s=1}^{C_{\alpha}}s=\frac{C_{\alpha}\,(C_{\alpha}+1)}{2}, \\
&\sum_{s\in\alpha}k_s^2=\sum_{s=1}^{C_{\alpha}}k_s^2=\sum_{s=1}^{C_{\alpha}}s^2=\frac{C_{\alpha}\,(C_{\alpha}+1)\,(2\,C_{\alpha}+1)}{6}, \\
&\sum_t \Theta(k_s-k_t)=k_s-1.
\end{split}
\label{identities}
\end{equation}
By replacing \eqref{identities} into \eqref{intermediate}, and rearranging some terms, we finally obtain
\begin{equation}
f(\vek{k}^{(\alpha)})=\Biggl(-\frac{1}{2}+\frac{1}{3\,N} \Biggr)\,C_{\alpha}+\frac{C_{\alpha}^2}{2}-\frac{C_{\alpha}^3}{3\,N}.
     \label{finalf}
\end{equation}
By plugging \eqref{finalf} into \eqref{component2} and rearranging some terms, we obtain
\begin{equation}
    \mathcal{N}_{\alpha}=\frac{C_{\alpha}}{N}-\frac{2}{3\,N^2}\,C_{\alpha}\,(C_{\alpha}+1).
    \label{nalpha}
\end{equation}
This represents the nestedness of a stepwise block $\alpha$ composed of $C_\alpha$ nodes.

We calculate now the in-block nestedness, $\mathcal{I}_0$ of a (disconnected) network composed of a set of disconnected stepwise blocks. This is readily obtained by summing the contributions $\mathcal{N}_\alpha$ -- given by Eq.~\eqref{nalpha} -- over all the blocks that compose the network.
In the limit scenario where all the blocks are small compared to the network ($C_{\alpha}\ll N$), we obtain $\mathcal{N}_{\alpha}\simeq C_{\alpha}/N$ and, as a consequence, $\mathcal{I}_0\simeq 1$.
In the general case, we obtain
\begin{equation}
    \mathcal{I}_0=1-\frac{2}{3\,N}-\frac{2}{3\,N^2}\sum_{\alpha}C_{\alpha}^2.
\end{equation}
In the case of equally-sized blocks, $C_{\alpha}=C=N/B$, we obtain Eq.~\eqref{unperturbed}. We verified numerically that this relation is correct.

\subsection{Proving the absence of a resolution limit}
\label{derivation2}

As detailed in Section~\ref{sec:analytic}, to prove the absence of a resolution limit, we need to calculate $\mathcal{I}_{single}$ and $\mathcal{I}_{pairs}$, and evaluate their difference.
To calculate $\mathcal{I}_{single}$, we perturb the perfectly in-block nested structure described above by connecting all the hubs of the $B$ blocks; each hub is now connected with two other hubs ($B$ inter-block links, in total) -- see Fig.~\ref{fig:fig1} for an illustration. Because of their links with the hubs of the two adjacent blocks, the hubs have degree $C_{\alpha}+2$. 
For simplicity, we assume that the blocks have the same size $C=N/B$; the hubs' degree is therefore $C+2=N/B+2$. 
In Eq.~\eqref{ibn2}, the $O_{st}/k_t$ term remains always equal to one if $k_s>k_t$ because internally, the blocks remain perfectly nested. 
The negative term receives now, for each block, an additional contribution given by the two extra link of each hub.
Therefore, the in-block nestedness of the network, $\mathcal{I}$, can be expressed as $\mathcal{I}=\mathcal{I}_0+\mathcal{I}_{int}$, where $\mathcal{I}_{int}$ is the "interaction" term that results from the edges that connect the hubs.
Overall, this extra term is
\begin{equation}
\mathcal{I}_{int}=-\frac{2}{N}\sum_{\alpha=1}^B \frac{1}{C_{\alpha}-1}\sum_{t\in\alpha}\frac{2}{N}\Theta(C+2-k_t)=-\frac{4\,B}{N^2},
\end{equation}
where we used the fact that there are $C_{\alpha}-1$ nodes of degree smaller than $C+2$ in each block (all the non-hub nodes, simply). Therefore, we obtain Eq.~\ref{Isingle}. We verified numerically that this relation is correct.

The challenge is now to calculate $\mathcal{I}_{pairs}$, i.e., the in-block nestedness for a partition where pairs of adjacent blocks are connected. 
For a given pair of blocks, $(\alpha_1,\alpha_2)$, we define the useful quantities:
\begin{equation}
    \begin{split}
        f(\vek{k}^{\alpha_{12}})&=\sum_{s,t\in\alpha_{12}}\Biggl(\frac{O_{st}}{k_t}-\frac{k_s}{N}\Biggr)\Theta(k_s-k_t), \\
        f_{11}(\vek{k}^{\alpha_{1}})&=\sum_{s,t\in\alpha_{1}}\Biggl(\frac{O_{st}}{k_t}-\frac{k_s}{N}\Biggr)\Theta(k_s-k_t),\\
        f_{12}(\vek{k}^{\alpha_{12}})&=\sum_{s\in\alpha_1,t\in\alpha_2}\Biggl(\frac{O_{st}}{k_t}-\frac{k_s}{N}\Biggr)\Theta(k_s-k_t).
    \end{split}
\end{equation}
We stress the fundamental difference among these quantities: $f(\vek{k}^{\alpha_{12}})$ is obtained from the contribution from all pairs of nodes that belong to the merged block $\alpha_{12}$; $f_{11}(\vek{k}^{\alpha_{1}})$ only receives contributions from pairs of nodes that belong to the same in-block nested block $\alpha_1$; $f_{12}(\vek{k}^{\alpha_{12}})$ only includes the contributions from pairs of nodes that belong to the same merged block $\alpha_{12}$, but different in-block nested blocks $\alpha_1$ and $\alpha_2$, respectively.
Based on symmetry with respect to permutations of the blocks, we obtain:
\begin{equation}
     \mathcal{I}_{pairs}=\frac{B}{2}\frac{2}{N}\frac{1}{2\,C-1}\,f(\vek{k}^{\alpha_{12}}).
     \label{Ipairsprov}
\end{equation}
Note that the block-size normalization factor is given by $1/(2\,C-1)$ and there is an overall factor $B/2$, which reflects the property that the partition comprises $B/2$ merged blocks which contain $2\,C$ nodes each.
For symmetry with respect to permutation of $\alpha_1$ and $\alpha_2$, we also have:
\begin{equation}
    f(\vek{k}^{\alpha_{12}})=2\,f_{11}(\vek{k}^{\alpha_{1}})+2\,f_{12}(\vek{k}^{\alpha_{12}}).
\end{equation}
By using this identity, we obtain
\begin{equation}
     \mathcal{I}_{pairs}=\frac{C-1}{2\,C-1}\,\mathcal{I}_{single}+\frac{2}{C\,(2\,C-1)}\,f_{12}(\vek{k}^{\alpha_{12}}),
\end{equation}
where we used the identity
\begin{equation}
    \mathcal{I}_{single}=\frac{2}{N}\,\frac{1}{C-1}\,f_{11}(\vek{k}^{\alpha_{1}}).
\end{equation}
In order to compare $\mathcal{I}^{pairs}$ against $\mathcal{I}^{single}$, the calculation of $f_{11}(\vek{k}^{\alpha_{1}})$ is left.
We obtain
\begin{equation}
    f_{12}(\vek{k}^{\alpha_{12}})=\sum_{t=1}^{C-1}\frac{1}{t}-\frac{(C-1)\,(C+2)}{N}-\sum_{s=1}^{C-1}\frac{s\,(s-1)}{N}
\end{equation}
The three terms on the r.h.s. have a clear interpretation. The first term is the positive contribution that comes from the overlap between the hub of block $\alpha_1$ and the $C-1$ non-hubs of block $\alpha_2$
The second term is the negative contribution that comes from the expected overlap between the hub of block $\alpha_1$ (with degree $C+2)$) and the $C-1$ non-hubs of block $\alpha_2$.
The third term is the negative contribution that comes from the expected overlap between the non-hubs of block $\alpha_1$ and the non-hubs of block $\alpha_2$; note that there is no overlap between the neighborhoods of the non-hubs of block $\alpha_1$ and the non-hubs of block $\alpha_2$.
By using again the identities \eqref{identities} and rearranging some terms, we obtain
\begin{equation}
    f_{12}(\vek{k}^{\alpha_{12}})=H_{C-1}-\frac{g(C)}{3\,N}
    \label{f12}
\end{equation}
where $H_{C-1}:=\sum_{t=1}^{C-1}t^{-1}$ denotes the $C-1$th harmonic number, and we defined the polynomial function $g(C):=(C-1)\,(C^2+C+6)$.
Note that the two terms in the r.h.s. represent the contribution of the observed and expected overlap between the nodes that belong to the two different original blocks that are joint together in the merged partition.
By plugging Eq.~\eqref{f12} into Eq.~\eqref{Ipairsprov}, we obtain
Eq.~\eqref{Ipairs} which, combined with Eq.~\ref{Isingle}, implies Eq.~\eqref{final}.


%

\end{document}